\RequirePackage[2020-02-02]{latexrelease}
\documentclass[onecolumn,amsmath,amssymb,10pt,aps]{revtex4}
  
\usepackage[utf8]{inputenc}
\usepackage[T1]{fontenc}

\usepackage{amsmath}
\usepackage{amsfonts}
\usepackage{graphicx}
\usepackage{yfonts}
\usepackage{color}
\usepackage[normalem]{ulem}
\usepackage{amsthm}
\usepackage{bm}
\usepackage{bbm}
\usepackage{mathtools}
\usepackage{array}
\usepackage{placeins}
\usepackage{enumitem}

\newcommand\bigforall{\mbox{\Large $\mathsurround=1pt\forall$}}

\def\<{\langle}
\def\>{\rangle}

\newcommand{\Tr}{\mathrm{Tr}}
\def\oper{{\mathchoice{\rm 1\mskip-4mu l}{\rm 1\mskip-4mu l}
{\rm 1\mskip-4.5mu l}{\rm 1\mskip-5mu l}}}
\DeclareMathAlphabet\mathbfcal{OMS}{cmsy}{b}{n}

\mathchardef\mhyphen="2D 

\newtheorem{Definition}{Definition}

\newtheorem{Proposition}{Proposition}
\newtheorem{Example}{Example}

\begin{document}


\title{Informationally overcomplete measurements from generalized equiangular tight frames}

\author{Katarzyna Siudzi\'{n}ska}
\affiliation{Institute of Physics, Faculty of Physics, Astronomy and Informatics, Nicolaus Copernicus University in Toru\'{n}, ul.~Grudzi\k{a}dzka 5, 87--100 Toru\'{n}, Poland}

\begin{abstract}
Informationally overcomplete measurements find important applications in quantum tomography and quantum state estimation. The most popular are maximal sets of mutually unbiased bases, for which trace relations between measurement operators are well known. In this paper, we introduce a more general class of informationally overcomplete POVMs that are generated by equiangular tight frames of arbitrary rank. This class provides a generalization of equiangular measurements to non-projective POVMs, which include rescaled mutually unbiased measurements and bases. We provide a method of their construction, analyze their symmetry properties, and provide examples for highly symmetric cases. In particular, we find a wide class of generalized equiangular measurements that are conical 2-designs, which allows us to derive the index of coincidence. Our results show benefits of considering a single informationally overcomplete measurement over informationally complete collections of POVMs.
\end{abstract}

\flushbottom

\maketitle

\thispagestyle{empty}

\section{Introduction}

Generalized quantum measurements are represented by positive, operator-valued measures (POVMs), which are positive operators $E_k$ whose sum $\sum_kE_k=\mathbb{I}_d$ is the identity operator. A celebrated example are symmetric, informationally complete (SIC) POVMs \cite{Renes} -- projective measurements characterized by strong symmetry conditions. Research on SIC POVMs is dictated by their wide experimental implementations \cite{Medendorp,Pimenta,Bent,Tavakoli,Tavakoli2,Smania}. Their applications include quantum state tomography \cite{Scott,ZhuEnglert,PetzRuppert}, quantum key distribution protocols \cite{Bouchard}, entanglement detection \cite{ESIC,EW-SIC,KalevBae}, and quantum state discrimination \cite{Brunner}. SIC POVMs are also analyzed in the context of non-locality \cite{OperationalSICs,Bene} and complex projective 2-designs \cite{Graydon,Junu1}. In recent years, much attention has been given to construct their generalizations. General SIC POVMs \cite{Gour,Yoshida} are operators of arbitrary rank. Semi-SIC POVMs \cite{semi-SIC} and equioverlapping measurements \cite{EOM22,EOM24,EOMq3} relax the condition of equal trace. Symmetric measurements \cite{SIC-MUB,SIC-MUB_general} are collections of POVMs with strong symmetry conditions.

A special property of SIC POVMs is that there are equal angles between all pairs of their vectors. Preserving the requirement of equal angles allows one to define equiangular measurements, which are closely related to other important mathematical objects: equiangular lines \cite{EAL,EAL2,EAL3,EAL4} and equiangular tight frames \cite{ETF1,ETF2}. Equiangular measurements have been used in detection and quantification of quantum entanglement \cite{EAM_App2,EAM_App3} with recent experimental applications in quantum key distribution \cite{EAM_App4} and quantum cryptography \cite{EAM_App1}. These measurements are projective but not informationally complete, which means that they do not allow for a full tomography of an arbitrary quantum state.

The main goal of this paper is to recover the property of informational completeness while also preserving the equiangularity of measurements. In our construction, we use families of complementary equiangular tight frames to define a generalized equiangular measurement. Its elements are in general non-projective and form an informationally overcomplete set. This means that they span the entire operator space, and some of the measurement operators are linearly dependent. Therefore, these POVMs provide a wide class of symmetric informationally overcomplete measurements, which find important applications e.g. in quantum state estimation \cite{IOC1,IOC2}, quantum tomography \cite{IOC3,IOC4,IOC5}, and quantum walks \cite{IOC6}. We also present a simple construction method based on the recently introduced generalized symmetric measurements \cite{SIC-MUB_general}.

In the following sections, we further analyze the properties of generalized equiangular measurements. In particular, we ask how much symmetry characteristic of the SIC POVMs can be recovered via clever manipulations of the associated parameters. Next, we propose a method of construction from the generalized symmetric measurements. We find three special classes of generalized equiangular measurements that are conical 2-designs. Importantly, two of these classes are constructed from the generalized symmetric measurements that themselves are not conical 2-designs. Finally, we derive the correspondence between the purity and index of coincidence for mixed quantum states. It turns out that it follows only for conical 2-designs. In the last section, we present a summary and a list of open problems that have arised during our research.

\section{Generalized equiangular	measurements}

In quantum mechanics, pure states correspond to collections of vectors (rays) $\{e^{i\theta}\psi;\,\theta\in\mathbb{R}\}$ in a complex Hilbert space $\mathcal{H}\simeq\mathbb{C}^d$. Two vectors $\psi$, $\phi$ are from the same ray if they differ only in the global phase $\theta$, so that $\phi=e^{i\theta}\psi$ for some angle $\theta$. If instead $\phi=c\psi$ for a complex number $c\in\mathbb{C}/\{0\}$, then form a line $\{c\psi;\,c\in\mathbb{C}\}$. A set $\{c_k\psi_k;\,k=1,\ldots,M\}$ contains equiangular lines if $|\<\psi_k|\psi_\ell\>|^2=c$ for all $k\neq\ell$. Equivalently, this can be reformulated in the language of rank-1 operators $P_k=a_k|\psi_k\>\<\psi_k|$ with $a_k=|c_k|^2$. Namely, $\{P_k;\,k=1,\ldots,M\}$ is a set of equiangular lines if $\Tr(P_kP_\ell)=c\Tr(P_k)\Tr(P_\ell)$. If additionally $\sum_{k=1}^MP_k=\gamma\mathbb{I}_d$ for $\gamma>0$, then $P_k$ form an equiangular tight frame \cite{Strohmer}. It is known that the number $M$ of elements $P_k$ is bounded by $d\leq M\leq d^2$, with $M=d^2$ corresponding to SIC POVMs \cite{Lemmens}. Equiangular tight frames are used to define measurement operators.

\begin{Definition}(\cite{EOM22})
A POVM $\{P_k;\,k=1,\ldots,M\}$ with the number of elements $d\leq M\leq d^2$ is an equiangular measurement if $P_k$ are rank-1 operators such that
\begin{equation}
\begin{split}
\Tr(P_k)&=a_k,\\
\Tr(P_kP_\ell)&=c\Tr(P_k)\Tr(P_\ell),\qquad k\neq\ell.
\end{split}
\end{equation}
\end{Definition}

From definition, it follows that the equiangular measurement operators are of equal trace and equal overlap;
\begin{equation}
a_k=\frac{d}{M},\qquad c=\frac{M-d}{d(M-1)}.
\end{equation}
Therefore, they are a special case of more general concepts: equioverlapping measurements \cite{EOM22,EOM24} and $(1,M)$-POVMs \cite{SIC-MUB}. As their number of elements is $M\leq d^2$, these POVMs are in general not informationally complete.

Trying to recover informational completeness, we consider collections of $N\geq 2$ equiangular measurements $\mathcal{E}_\alpha=\{P_{\alpha,k};\,k=1,\ldots,M_\alpha\}$ whose total number of elements satisfies
\begin{equation}\label{MN}
\sum_{\alpha=1}^NM_\alpha\geq d^2+N-1.
\end{equation}
However, as it turns out, this is not enough to guarantee $d^2$ linearly independent operators among $P_{\alpha,k}$. As an example, take the qubit case ($d=2$), where it is known how to construct projective POVMs for the whole range of $M=2,3,4$. Let us take $N=2$ equiangular measurements with $M_1=M_2=3$:
\begin{equation}
P_{1,1}=\frac 13 \begin{pmatrix} 1 & 1 \\ 1 & 1 \end{pmatrix},\qquad
P_{1,2}=\frac 16 \begin{pmatrix} 2 & -1+i\sqrt{3} \\ -1-i\sqrt{3} & 2 \end{pmatrix},\qquad
P_{1,3}=\frac 16 \begin{pmatrix} 2 & -1-i\sqrt{3} \\ -1+i\sqrt{3} & 2 \end{pmatrix},
\end{equation}
and
\begin{equation}
P_{2,1}=\frac 13 \begin{pmatrix} 1 & -i \\ i & 1 \end{pmatrix},\qquad
P_{2,2}=\frac 16 \begin{pmatrix} 2 & i+\sqrt{3} \\ -i+\sqrt{3} & 2  \end{pmatrix},\qquad
P_{2,3}=\frac 16 \begin{pmatrix} 2 & i-\sqrt{3} \\ -i-\sqrt{3} & 2
\end{pmatrix}.
\end{equation}
Observe that, despite the total number of operators exceeding $d^2+N-1=5$, $P_{\alpha,k}$ indeed do not span the entire space of Hermitian operators on $\mathbb{C}^2$. Instead, this set is overcomplete on a subspace spanned by $\{\mathbb{I}_2,\sigma_1,\sigma_2\}$. Both measurements satisfy the symmetry conditions
\begin{equation}
\Tr P_{\alpha,k}P_{\alpha,\ell}=\frac 19,
\end{equation}
but some of the symmetry for elements from different POVMs is lost. In this example,
\begin{equation}
\Tr P_{1,k}P_{2,k}=\frac 29,\qquad \Tr P_{1,k}P_{2,k+1}=\frac{2+\sqrt{3}}{9},
\qquad \Tr P_{1,k}P_{2,k+2}=\frac{2-\sqrt{3}}{9},
\end{equation}
where $k+1$ and $k+2$ are calculated modulo 2. This pair of equiangular measurements can be completed by e.g. the projective POVM $P_{3,k}=|k\>\<k|$, for which
\begin{equation}
\Tr P_{3,k}P_{1,m}=\Tr P_{3,k}P_{2,n}=\frac 13.
\end{equation}

Evidently, more restrictive constraints are needed. Let us proceed with the following steps.
\begin{enumerate}[label=(\arabic*)]
\item Consider collections of equiangular tight frames, so that $\sum_{k=1}^{M_\alpha}P_{\alpha,k}=\gamma_\alpha\mathbb{I}_d$ with $\gamma_\alpha>0$. Moreover, the collection $\{P_{\alpha,k};\,k=1,\ldots,M_\alpha;\,\alpha=1,\ldots,N\}$ itself is a POVM.
\item Impose a complementarity condition \cite{PetzRuppert} $\Tr(P_{\alpha,k}P_{\beta,\ell})=f\Tr(P_{\alpha,k})\Tr(P_{\beta,\ell})$ on the elements from different equiangular tight frames ($\alpha\neq\beta$).
\item Although not strictly necessary at this point, we generalize the equiangular tight frames to also allow for non-projective positive operators $P_{\alpha,k}/a_k$. This way, the lower bound on the number of elements becomes $M_\alpha\geq 2$ instead of $M_\alpha\geq d$.
\end{enumerate}
In other words, we introduce a generalization of equiangular measurements by constructing a POVM from $N\geq 2$ arbitrary rank equivalents of equiangular tight frames. This results in the following definition.

\begin{Definition}\label{geam}
A collection of $N$ generalized equiangular lines $\{P_{\alpha,k};\,k=1,\ldots,M_\alpha;\,\alpha=1,\ldots,N\}$ such that $\sum_{k=1}^{M_\alpha}P_{\alpha,k}=\gamma_\alpha\mathbb{I}_d$ and $\sum_{\alpha=1}^N\gamma_\alpha=1$ is a generalized equiangular measurement if
\begin{equation}
\begin{split}
\Tr(P_{\alpha,k})&=a_\alpha,\\
\Tr(P_{\alpha,k}^2)&=b_\alpha \Tr(P_{\alpha,k})^2,\\
\Tr(P_{\alpha,k}P_{\alpha,\ell})&=c_\alpha
\Tr(P_{\alpha,k})\Tr(P_{\alpha,\ell}),\qquad k\neq\ell,\\
\Tr(P_{\alpha,k}P_{\beta,\ell})&=
f\Tr(P_{\alpha,k})\Tr(P_{\beta,\ell}),\qquad \alpha\neq\beta.
\end{split}
\end{equation}
\end{Definition}

From Definition 2, we find the coefficients that characterize the generalized equiangular measurements,
\begin{equation}
a_\alpha=\frac{d\gamma_\alpha}{M_\alpha},\qquad c_\alpha=\frac{M_\alpha-db_\alpha}{d(M_\alpha-1)},
\qquad f=\frac 1d,
\end{equation}
and the admissible range for $b_\alpha$,
\begin{equation}\label{ba}
\frac 1d <b_\alpha\leq\frac 1d \min\{d,M_\alpha\}.
\end{equation}
Equiangular measurements from Definition 1 are recovered for $N=\gamma_\alpha=b_\alpha=1$. In Appendix \ref{AppA}, we prove that the generalized equiangular measurement is informationally overcomplete if and only if the equality in eq. (\ref{MN}) holds, and informational completeness is achieved only for $N=1$. The highest admissible value elements is $2(d^2-1)$, and it corresponds to the choice $N=d^2-1$, $M_\alpha=2$.

Observe that one parameter, $f=1/d$, is always constant, independent on the indices $\alpha\neq\beta$. Moreover, if $\gamma_\alpha=aM_\alpha/d$ with $a=d/(d^2+N-1)$, then the measurement elements are of equal trace $a_\alpha=a$. Independently of the values of $a_\alpha$, one of the remaining coefficients can be set to be line-independent.
\begin{itemize}
\item If $1/d<b\leq\min\{1,M_\alpha/d\}$ is a non-empty set, then there exists a generalized equiangular measurement for which $b_\alpha=b$. However,  $c_\alpha\neq c_\beta$ as long as $M_\alpha\neq M_\beta$.
\item If $\max\{0,\frac{M_\alpha-d}{d(M_\alpha-1)}\}<c\leq 1/d$ is a non-empty set, then there exists a generalized equiangular measurement such that $c_\alpha=c$. In this case, $b_\alpha\neq b_\beta$ whenever $M_\alpha\neq M_\beta$.
\end{itemize}
Therefore, there can be up to three parameters independent on $\alpha$: $(a,b,c_\alpha,f)$ or $(a,b_\alpha,c,f)$. However, there is one more important class.

\begin{Proposition}\label{design1}
If the set
\begin{equation}
0<R\leq\min\left\{\frac{M_\alpha}{d},\frac{M_\alpha(d-1)}{d(M_\alpha-1)}\right\}
\end{equation}
is non-empty, then there exists a generalized equiangular measurement with $b_\alpha-c_\alpha=R$.
\end{Proposition}

Note that a constant difference between $b_\alpha=1/d+R(M_\alpha-1)/M_\alpha$ and $c_\alpha=1/d-R/M_\alpha$ is possible despite the fact that both depend on the choice of a line $\alpha$.

\section{Relation to generalized symmetric measurements}

The generalized symmetric measurements have been introduced as collections of unequinumerous POVMs that generalize the notion of $(N,M)$-POVMs \cite{SIC-MUB}. They can be defined in any finite dimension.

\begin{Definition}(\cite{SIC-MUB_general})
Generalized symmetric measurements are collections of $N$ POVMs $\{E_{\alpha,k};\, k=1,\ldots,M_\alpha\}$ satisfying additional symmetry conditions
\begin{equation}\label{MM}
\begin{split}
\Tr (E_{\alpha,k})&=w_\alpha=\frac{d}{M_\alpha},\\
\Tr (E_{\alpha,k}^2)&=x_\alpha,\\
\Tr (E_{\alpha,k}E_{\alpha,\ell})&=y_\alpha=\frac{d-M_\alpha x_\alpha}{M_\alpha(M_\alpha-1)},\qquad \ell\neq k,\\
\Tr (E_{\alpha,k}E_{\beta,\ell})&=z_{\alpha\beta}=\frac{d}{M_\alpha M_\beta},\qquad\beta\neq\alpha,
\end{split}
\end{equation}
where
\begin{equation}
\frac{d}{M_\alpha^2}<x_\alpha\leq \min\left\{\frac{d^2}{M_\alpha^2},\frac {d}{M_\alpha}\right\}.
\end{equation}
\end{Definition}

For informationally complete measurements, where  $\sum_{\alpha=1}^NM_\alpha=d^2+N-1$, four construction methods are known. Each uses different linear combinations of Hermitian orthonormal operator bases $\{G_0=\mathbb{I}_d/\sqrt{d},G_{\alpha,k};\,k=1,\ldots,M_\alpha;\,\alpha=1,\ldots,N\}$, where $G_{\alpha,k}$ are traceless operators. The conditions for every construction to produce distinct sets of measurement operators have also been analyzed. After fixing the construction method, the correspondence between $E_{\alpha,k}$ and $G_{\alpha,k}$ is one-to-one \cite{SIC-MUB_general}.

There is a close relation between the generalized symmetric measurements and the generalized equiangular measurements. Obviously, if all $\gamma_\alpha=1/N$, then $\{E_{\alpha,k}\}=\{NP_{\alpha,k}\}$. Actually, for any other choice of $\gamma_\alpha$, there is a linear correspondence between $E_{\alpha,k}$ and $P_{\alpha,k}$. This property proves useful in constructing the generalized equiangular measurements.

\begin{Proposition}
A generalized equiangular measurement $P_{\alpha,k}$ is constructed from the generalized symmetric measurements $E_{\alpha,k}$ via
\begin{equation}
P_{\alpha,k}=\gamma_\alpha E_{\alpha,k},
\end{equation}
where $\gamma_\alpha$ is a probability distribution.
\end{Proposition}

The correspondence between $(a_\alpha,b_\alpha,c_\alpha,f)$ and $(w_\alpha,x_\alpha,y_\alpha,z_{\alpha\beta})$ reads
\begin{equation}
a_\alpha=\gamma_\alpha w_\alpha,\qquad
b_\alpha=\frac{x_\alpha}{w_\alpha^2},\qquad
c_\alpha=\frac{y_\alpha}{w_\alpha^2},\qquad
f=\frac{z_{\alpha\beta}}{w_\alpha w_\beta}.
\end{equation}
Note that $P_{\alpha,k}$ form an informationally overcomplete set if and only if they are constructed from informationally complete $E_{\alpha,k}$ and $\gamma_\alpha\neq 0$.

\begin{Example}
A popular example of the generalized equiangular measurement is a POVM constructed from $d+1$ MUBs by taking $\gamma_\alpha=1/(d+1)$. In this case, one has
\begin{equation}
a_\alpha=\frac{1}{d+1},\qquad
b_\alpha=1,\qquad
c_\alpha=0,\qquad
f=\frac 1d.
\end{equation}
\end{Example}

\begin{Example}
In $d=2$, take $N=2$ informationally complete generalized symmetric measurements with $M_1=2$ and $M_2=3$ elements, respectively. In particular, the 2-elemental POVM is the von Neumann measurement,
\begin{equation}
E_{1,1}=\begin{pmatrix} 1 & 0 \\ 0 & 0
\end{pmatrix},\qquad 
E_{1,2}=\begin{pmatrix} 0 & 0 \\ 0 & 1
\end{pmatrix},
\end{equation}
and the 3-elemental POVM is projective,
\begin{equation}
E_{2,1}=\frac 13 \begin{pmatrix} 1 & -i \\ i & 1 \end{pmatrix},\qquad
E_{2,2}=\frac 16 \begin{pmatrix} 2 & i+\sqrt{3} \\ -i+\sqrt{3} & 2  \end{pmatrix},\qquad
E_{2,3}=\frac 16 \begin{pmatrix} 2 & i-\sqrt{3} \\ -i-\sqrt{3} & 2
\end{pmatrix}.
\end{equation}
Now, a generalized equiangular measurement is obtained via $P_{1,k}=(1-\gamma)E_{1,k}$ and $P_{2,k}=2\gamma E_{2,k}/3$, where $0<\gamma<1$. If one fixes $\gamma=3/5$, then all the elements $P_{\alpha,k}$ and their squares are of equal trace,
\begin{equation}
\Tr (P_{\alpha,k})=\frac 25,\qquad
\Tr (P_{\alpha,k}^2)=\frac{4}{25},
\end{equation}
whereas the trace relations between its pairs depend on the choice of equiangular lines,
\begin{equation}
\Tr (P_{1,1}P_{1,2})=0,\qquad
\Tr (P_{2,1}P_{2,2})=\frac{1}{25},\qquad
\Tr (P_{1,k}P_{2,\ell})=\frac{2}{25}.
\end{equation}
\end{Example}

To summarize, a generalized equiangular measurement $P_{\alpha,k}$ are constructed from the generalized symmetric measurements $E_{\alpha,k}$ through a simple rescaling. Even though this construction may seem trivial, it allows for more freedom in parameter manipulations compared to the generalized symmetric measurements. For example, equal trace of elements is only guaranteed for the symmetric measurements ($M_\alpha=M$), whereas the elements of a generalized equiangular measurement are of equal trace for $\gamma_\alpha=aM_\alpha/d$ (even if $M_\alpha\neq M_\beta$). This property makes it possible for $P_{\alpha,k}$ to present interesting behaviour inaccessible to $E_{\alpha,k}$. For one, the generalized symmetric measurements allow for at most one constant parameter, $x_\alpha=x$ or $y_\alpha=y$, in the most general case where $M_\alpha\neq M_\beta$. One more is possible for $N=2$, which is $z_{\alpha\beta}=z$ \cite{SIC-MUB_general}. In the case of the generalized equiangular measurements, the additional degrees of freedom in the form of $\gamma_\alpha$ can be manipulated to guarantee a higher number of constant traces.

\begin{Proposition}\label{design2}
Consider the generalized symmetric measurements characterized by
\begin{equation}
x_\alpha=\frac{d}{M_\alpha}\frac{1}{1+\eta(M_\alpha-1)},\qquad y_\alpha=\eta x_\alpha,\qquad
\max\left\{0,\frac{M_\alpha-d}{d(M_\alpha-1)}\right\}\leq\eta
<1,
\end{equation}
and the probability distribution $\gamma_\alpha=\sqrt{[1+\eta(M_\alpha-1)]BM_\alpha/d}$.
The corresponding generalized equiangular measurement satisfies
\begin{equation}
\begin{split}
\Tr(P_{\alpha,k})&=a_\alpha=\frac{d\gamma_\alpha}{M_\alpha},\\
\Tr(P_{\alpha,k}^2)&=B,\\
\Tr(P_{\alpha,k}P_{\alpha,\ell})&=C=\eta B,\qquad k\neq\ell,\\
\Tr(P_{\alpha,k}P_{\beta,\ell})&=F_{\alpha\beta}=
\frac{d\gamma_\alpha\gamma_\beta}{M_\alpha M_\beta},\qquad \alpha\neq\beta.
\end{split}
\end{equation}
Moreover, $a_\alpha=a_\beta$ if and only if $M_\alpha=M_\beta$, whereas $F_{\alpha\beta}=F$ also for $N=2$.
\end{Proposition}

Interestingly, in the case of the generalized equiangular measurement, it is possible to manipulate the coefficients in such a way that both $\Tr(P_{\alpha,k}^2)$ and $\Tr(P_{\alpha,k}P_{\alpha,\ell})$ become $\alpha$-independent. This behavior was not observed for the generalized symmetric measurements. Therefore, by allowing for overcompleteness of POVMs, one can recover more symmetries than for informationally complete POVMs.

\begin{Example}
A good example of a highly symmetric generalized equiangular measurement from Proposition \ref{design2} is
\begin{equation}
P_{1,1}=\frac{3-\sqrt{5}}{8}
\begin{pmatrix}
\sqrt{5}-\sqrt{3} & 0 \\ 0 & \sqrt{5}+\sqrt{3}
\end{pmatrix},\qquad
P_{1,2}=\frac{3-\sqrt{5}}{8}
\begin{pmatrix}
\sqrt{5}+\sqrt{3} & 0 \\ 0 & \sqrt{5}-\sqrt{3}
\end{pmatrix},
\end{equation}
\begin{equation}
P_{2,1}=\frac{3-\sqrt{5}}{8}
\begin{pmatrix}
2 & q \\ \overline{q} & 2
\end{pmatrix},\qquad
P_{2,2}=\frac{3-\sqrt{5}}{8}
\begin{pmatrix}
2 & -i\overline{q} \\ iq & 2
\end{pmatrix},\qquad
P_{2,3}=\frac{3-\sqrt{5}}{4\sqrt{2}}
\begin{pmatrix}
\sqrt{2} & 1-i \\ 1+i & \sqrt{2}
\end{pmatrix},
\end{equation}
\begin{equation}
q=-(2+\sqrt{3}+i)\sqrt{2-\sqrt{3}},
\end{equation}
for which $\gamma_1=(3-\sqrt{5})\sqrt{5}/4$ and $\gamma_2=1-\gamma_1$.
Indeed, there is no $\alpha$-dependence in the inner products between pairs,
\begin{equation}
\begin{split}
\Tr(P_{\alpha,k}^2)&=B=\frac 12 (7-3\sqrt{5}),\\
\Tr(P_{\alpha,k}P_{\alpha,\ell})&=C=\frac 18 (7-3\sqrt{5}),\qquad k\neq\ell,\\
\Tr(P_{\alpha,k}P_{\beta,\ell})&=F=\frac {\sqrt{5}}{8} (7-3\sqrt{5}),\qquad \alpha\neq\beta,
\end{split}
\end{equation}
even though the elements themselves are not of equal trace,
\begin{equation}
\Tr(P_{1,1})=\Tr(P_{1,2})=a_1=\frac{\sqrt{5}}{4}(3-\sqrt{5}),\qquad
\Tr(P_{2,1})=\Tr(P_{2,2})=\Tr(P_{2,3})=a_2=\frac 12 (3-\sqrt{5}).
\end{equation}
\end{Example}

\section{Conical designs}

Conical designs are generalizations of complex projective designs to positive operators. From definition, a family of positive operators $E_k$ is a conical $t$-design if $E_k^{\otimes t}$ commutes with a $t$-th tensor product $U^{\otimes t}$ of any unitary operator $U$ \cite{Graydon,Graydon2}. Most interest has been given to the special case with $t=2$. Equivalently, $E_k$ form a conical 2-design if and only if
\begin{equation}\label{con}
\sum_kE_k\otimes E_k=
\kappa_+\mathbb{I}_d\otimes\mathbb{I}_d+\kappa_-\mathbb{F}_d
\end{equation}
for $\kappa_+\geq\kappa_->0$ and the flip operator $\mathbb{F}_d=\sum_{m,n=1}^d|m\>\<n|\otimes|n\>\<m|$ \cite{Graydon}.
Examples include general SIC POVMs, mutually unbiased measurements \cite{Wang}, $(N,M)$-POVMs \cite{SICMUB_design,SICMUB_channels}, and a class of generalized symmetric measurements \cite{SIC-MUB_general}. We show that some of the generalized equiangular measurements are conical 2-designs, as well.

\begin{Proposition}\label{c2d}
The generalized equiangular measurement characterized by $a_\alpha^2(b_\alpha-c_\alpha)=S$ such that
\begin{equation}\label{Srange}
0<S\leq \min\left\{\frac{d\gamma_\alpha^2}{M_\alpha},\frac{d-1}{M_\alpha-1}\frac{d\gamma_\alpha^2}{M_\alpha}\right\}
\end{equation}
is a conical 2-design with
\begin{equation}\label{kappas2}
\kappa_+=\mu-\frac{S}{d},\qquad\kappa_-=S,
\end{equation}
where $\mu=(1/d)\sum_{\alpha=1}^Na_\alpha\gamma_\alpha$.
\end{Proposition}

The proof is presented in Appendix \ref{AppB}. Examples of conical 2-designs include three special classes:
\begin{enumerate}[label=(\it\roman*)]
\item $P_{\alpha,k}$ from Proposition \ref{design1}, for which $b_\alpha-c_\alpha=R$, but only under the additional condition that $a_\alpha=a=d/(d^2+N-1)$. In this case, $S=a^2R$.
\item $P_{\alpha,k}$ from Proposition \ref{design2}, for which $a_\alpha^2b_\alpha=B$ and $a_\alpha^2c_\alpha=C$. Then, the constant $S=B-C$.
\item $P_{\alpha,k}$ constructed from $\gamma_\alpha=1/N$ and $E_{\alpha,k}$ that are conical 2-designs -- that is, they are characterized by $x_\alpha-y_\alpha=r$ with \cite{SIC-MUB_general}
\begin{equation}
0<r\leq\min\left\{\frac{d}{M_\alpha},\frac{d(d-1)}{M_\alpha(M_\alpha-1)}\right\}.
\end{equation}
The parameters associated with the resulting generalized equiangular measurements read
\begin{equation}\label{design3}
a_\alpha=\frac{d}{NM_\alpha},\qquad b_\alpha=\frac{d+rM_\alpha(M_\alpha-1)}{d^2},\qquad c_\alpha=\frac{d-rM_\alpha}{d^2},
\end{equation}
so that $S=a_\alpha^2(b_\alpha-c_\alpha)=r/N^2$.
\end{enumerate}
Note that $(i)$ and $(ii)$ are conical 2-designs even though they are constructed from the generalized symmetric measurements that are not conical 2-designs.

\section{Index of coincidence}

A density operator that represents a mixed quantum state can be expanded into
\begin{equation}\label{rho}
\rho=\sum_{\alpha=1}^N\sum_{k=1}^{M_\alpha}p_{\alpha,k}Q_{\alpha,k}.
\end{equation}
Now, $\{p_{\alpha,k}=\Tr(\rho P_{\alpha,k});\,k=1,\ldots,M_\alpha;\,\alpha=1,\ldots,N\}$ is a probability distribution, whereas $Q_{\alpha,k}$ is a dual frame to $P_{\alpha,k}$. Due to overcompleteness of the generalized equiangular measurements, the choice of a dual frame is not unique. However, for our purposes, we distinguish a special frame: the one that can be constructed from
\begin{equation}\label{Fak}
F_{\alpha,k}=\frac{1}{x_\alpha-y_\alpha}\left[E_{\alpha,k}-\frac 1d \mathbb{I}_d\left(w_\alpha-\frac{x_\alpha-y_\alpha}{N}\right)\right],
\end{equation}
which is the frame dual to $E_{\alpha,k}$ \cite{SIC-MUB_general}. Namely, if we construct $P_{\alpha,k}=\gamma_\alpha E_{\alpha,k}$, as in Proposition 2, then
\begin{equation}\label{Qak}
Q_{\alpha,k}=\frac{1}{\gamma_\alpha}F_{\alpha,k}
=\frac{1}{a_\alpha^2(b_\alpha-c_\alpha)}\left[P_{\alpha,k}-\frac 1d \mathbb{I}_d\left(a_\alpha-\frac{a_\alpha^2(b_\alpha-c_\alpha)}{N\gamma_\alpha}\right)\right].
\end{equation}

Using the probability distribution $p_{\alpha,k}$ from eq. (\ref{rho}), one defines the index of coincidence \cite{Rastegin5}
\begin{equation}
\mathcal{C}=\sum_{\alpha=1}^N\sum_{k=1}^{M_\alpha}p_{\alpha,k}^2.
\end{equation}
As it tutns out, it can be analytically derived only for the generalized equiangular measurements that are conical 2-designs.

\begin{Proposition}\label{prop}
If $P_{\alpha,k}$ is a conical 2-design from Proposition \ref{c2d}, for which $a_\alpha^2(b_\alpha-c_\alpha)=S$, then its index of coincidence
\begin{equation}\label{IOC}
\mathcal{C}=S\left(\Tr\rho^2-\frac 1d\right)+\mu,
\end{equation}
where $\mu=(1/d)\sum_{\alpha=1}^Na_\alpha\gamma_\alpha$.
\end{Proposition}

For the proof, see Appendix \ref{AppC}. Note that $\Tr(\rho^2)\leq 1$, so the index of coincidence is upper bounded by
\begin{equation}
\mathcal{C}_{\max}=\frac{d-1}{d}S+\mu,
\end{equation}
with the equality reached for pure states.

\begin{Example}
Consider the three special classes of the generalized equiangular measurements that are conical 2-designs.
\begin{enumerate}[label=(\roman*)]
\item For $P_{\alpha,k}$ from Proposition \ref{design1} with $a_\alpha=a=d/(d^2-N+1)$, one has $S=a^2R$ and $\mu=a/d$. Therefore,
\begin{equation}
\mathcal{C}_{\max}=\frac ad [aR(d-1)+1].
\end{equation}
\item For $P_{\alpha,k}$ from Proposition \ref{design2}, the constants $S=(1-\eta)B$ and $\mu=[\eta(d^2-1)+N]B/d$. Then,
\begin{equation}
\mathcal{C}_{\max}=\frac Bd [(d-1)(d\eta+1)+1].
\end{equation}
\item For $P_{\alpha,k}$ constructed from conical 2-designs $E_{\alpha,k}$, it follows that $S=r/N^2$, $\mu=N^{-2}\sum_{\alpha=1}^NM_\alpha^{-1}$, and hence
\begin{equation}
\mathcal{C}_{\max}=\frac{1}{N^2} \left[\frac{d-1}{d}r+
\sum_{\alpha=1}^N\frac{1}{M_\alpha}\right].
\end{equation}
\end{enumerate}
\end{Example}

\section{Conclusions}

In this paper, we introduce a generalization of equiangular measurements by constructing a POVM from generalized equiangular tight frames that are complementary to one another. This results in informationally overcomplete POVMs, for which we present a method of construction from the generalized symmetric measurements. They provide an interesting and non-trivial generalization of informationally overcomplete POVMs obtained via rescaling complete sets of MUBs. Next, we propose several special classes that are characterized by high symmetries between its elements. In particular, we present an examaple of a generalized equiangular measurement with three out of four defining parameters being independent of the measurement operator indices. Then, we find a wide class of measurements that are conical 2-designs. For this class, we also derive an analytical expression for the index of coincidence.

Our results show that a single informationally overcomplete POVM can manifest interesting behaviors that are absent for an informationally complete collection of POVMs. Therefore, it may prove beneficial to analyze other classes, beyond the generalization of equiangular measurements. Furthermore, it is worth exploring whether there exists measurements that are not conical 2-designs but allow for a linear relation between the purity of states and the index of coincidence. If not, this could point us to the reason why these two properties seem to be so closely related. Other open questions include possible applications in quantum state estimation and quantum tomography, where informationally overcomplete measurements have already found their uses.

\section{Acknowledgements}

This research was funded in whole or in part by the National Science Centre, Poland, Grant number 2021/43/D/ST2/00102. For the purpose of Open Access, the author has applied a CC-BY public copyright licence to any Author Accepted Manuscript (AAM) version arising from this submission.

\appendix

\section{Informational overcompleteness}\label{AppA}

A POVM is informationally complete or overcomplete if and only if it consists in $d^2$ linearly independent operators. The elements of the generalized equiangular measurement associated with the same generalized equiangular line $\alpha$ are not all linearly independent due to the constraints $\sum_{k=1}^{M_\alpha}P_{\alpha,k}=\gamma_\alpha\mathbb{I}_d$. Therefore, let us take $M_\alpha-1$ elements from each line and define
\begin{equation}
\mathbb{P}=\sum_{\alpha=1}^N\sum_{k=1}^{M_\alpha-1}r_{\alpha,k}P_{\alpha,k}.
\end{equation}
Now, if $\mathbb{P}=0$ impies that $r_{\alpha,k}=0$, then $\{P_{\alpha,k};\,k=1,\ldots,M_\alpha-1;\,\alpha=1,\ldots,N\}$ is a linearly independent set. In what follows, we prove that this is indeed true. First, observe that
\begin{equation}
\Tr(\mathbb{P})=\sum_{\alpha=1}^N\sum_{k=1}^{M_\alpha-1}r_{\alpha,k}=0
\end{equation}
and, for any element $P_{\alpha,k}$ with $k=1,\ldots,M_\alpha-1$,
\begin{equation}\label{22}
\Tr(\mathbb{P}P_{\alpha,k})=(b_\alpha-c_\alpha)a_\alpha^2r_{\alpha,k}
+a_\alpha^2(c_\alpha-f)\sum_{\ell=1}^{M_\alpha-1}r_{\alpha,\ell}=0.
\end{equation}
Taking the sum over $k=1,\ldots,M_\alpha-1$, one gets
\begin{equation}
\sum_{k=1}^{M_\alpha-1}\Tr(\mathbb{P}P_{\alpha,k})=a_\alpha^2[b_\alpha-c_\alpha+
M_\alpha(c_\alpha-f)]\sum_{k=1}^{M_\alpha-1}r_{\alpha,k}=0.
\end{equation}
This means that $\sum_{k=1}^{M_\alpha-1}r_{\alpha,k}=0$ because $b_\alpha\neq c_\alpha\neq f$. Substituting these results into eq. (\ref{22}), we show that the only solution of $\mathbb{P}=0$ is $r_{\alpha,k}=0$. The total number of elements $\sum_{\alpha=1}^NM_\alpha=d^2+N-1\geq d^2$, where the equality only holds for $N=1$. Hence, the generalized equiangular measurements are informationally complete for $N=1$ and overcomplete otherwise.

\section{Conical designs}\label{AppB}

From ref. \cite{SIC-MUB_general}, the generalized equiangular measurement $\{P_{\alpha,k}\}$ is a conical 2-design
\begin{equation}
\sum_{\alpha=1}^N\sum_{k=1}^{M_\alpha}P_{\alpha,k}\otimes P_{\alpha,k}
=\kappa_+\mathbb{I}_d\otimes\mathbb{I}_d+\kappa_-\mathbb{F}_d
\end{equation}
if and only if it defines a linear map
\begin{equation}
\Phi[X]=\sum_{\alpha=1}^N\sum_{k=1}^{M_\alpha}P_{\alpha,k}\Tr(XP_{\alpha,k})
\end{equation}
such that
\begin{equation}\label{suma}
\Phi=\kappa_-\oper+\kappa_+d\Phi_0,\qquad \kappa_+\geq\kappa_->0,
\end{equation}
with the maximally depolarizing channel $\Phi_0[X]=\mathbb{I}_d\Tr(X)/d$ and the identity operator $\oper$. To derive the conditions under which eq. (\ref{suma}) holds, let us compute
\begin{equation}
\Tr(\Phi[X]P_{\alpha,k})
=a_\alpha^2(b_\alpha-c_\alpha)\Tr(P_{\alpha,k}X)
+\left[a_\alpha^2\gamma_\alpha(c_\alpha-f)
+a_\alpha f\sum_{\beta=1}^Na_\beta\gamma_\beta\right]
\Tr X.
\end{equation}
Knowing that $c_\alpha-f=-(b_\alpha-c_\alpha)/M_\alpha$, this can be equivalently rewritten into
\begin{equation}\label{trace2}
\Tr\left\{\left(\Phi-\left[a_\alpha^2(b_\alpha-c_\alpha)\oper
+\left(-a_\alpha^2(b_\alpha-c_\alpha)+\sum_{\beta=1}^Na_\beta\gamma_\beta\right)\Phi_0
\right]\right)[X]P_{\alpha,k}\right\}=0.
\end{equation}
The above relation has to be satisfied for any measurement operator $P_{\alpha,k}$ and Hermitian $X$. Therefore, the coefficients that multiply $\oper$ and $\Phi_0$ are required to be independent on the index $\alpha$, which is the case for $a_\alpha^2(b_\alpha-c_\alpha)=S$. Then, $\Phi$ is of the form given in eq. (\ref{suma}) with $\kappa_\pm$ from eq. (\ref{kappas2}).

The admissible range of $S$ in eq. (\ref{Srange}) follows directly from the range of $b_\alpha$ in eq. (\ref{ba}). Moreover, from eq. (\ref{Srange}), it follows that $\kappa_->0$,
\begin{equation}
\bigforall_\alpha\quad S\leq\frac{d\gamma_\alpha^2}{M_\alpha}\qquad{\rm and}
\qquad \bigforall_\alpha\quad S
\leq\frac{d-1}{M_\alpha-1}\frac{d\gamma_\alpha^2}{M_\alpha}.
\end{equation}
By taking the sum over all $\alpha=1,\ldots,N$, we arrive at
\begin{equation}
S\leq\frac{d\mu}{\max\{N,d+1\}},
\end{equation}
which clearly shows that
\begin{equation}
\kappa_+-\kappa_-=\mu-\frac{d+1}{d}S\geq 0.
\end{equation}
Hence, all generalized equiangular measurements for which $a_\alpha^2(b_\alpha-c_\alpha)=S$ are conical 2-designs.

\section{Index of coincidence}\label{AppC}

First, let us recall that any mixed state $\rho$ admits the following decomposition,
\begin{equation}
\rho=\sum_{\alpha=1}^N\sum_{k=1}^{M_\alpha}p_{\alpha,k}Q_{\alpha,k},
\end{equation}
where $p_{\alpha,k}$ is a probability distribution. The operators $Q_{\alpha,k}$ denote the dual frame defined in eq. (\ref{Qak}), and they obey the trace relations
\begin{equation}
\Tr(Q_{\alpha,k}Q_{\beta,\ell})=\frac{1}{a_\alpha^2 
(b_\alpha-c_\alpha) a_\beta^2(b_\beta-c_\beta)}
[\Tr(P_{\alpha,k}P_{\beta,\ell})-fa_\alpha a_\beta]
+\frac{f}{N^2\gamma_\alpha\gamma_\beta},
\end{equation}
where
\begin{equation}
\Tr(P_{\alpha,k}P_{\beta,\ell})=\delta_{\alpha\beta}\delta_{k\ell}a_\alpha^2(b_\alpha-c_\alpha)
+\delta_{\alpha\beta}a_\alpha^2(c_\alpha-f)+fa_\alpha a_\beta.
\end{equation}
Now, remembering that $\sum_{k=1}^{M_\alpha}p_{\alpha,k}=\gamma_\alpha$, we calculate the purity
\begin{equation}\label{rho2}
\Tr\rho^2=\frac 1d +\sum_{\alpha=1}^N\frac{1}{a_\alpha^2(b_\alpha-c_\alpha)}\left(\sum_{k=1}^{M_\alpha}p_{\alpha,k}^2
-\frac{a_\alpha\gamma_\alpha}{d}\right).
\end{equation}
Note that, in general, there is no clear correspondence between $\Tr\rho^2$ and the index of coincidence $\mathcal{C}$. However, under the assumption that $a_\alpha^2(b_\alpha-c_\alpha)=S$, one recovers
\begin{equation}
\Tr\rho^2=\frac 1d +\frac{\mathcal{C}-\mu}{S},
\end{equation}
which is equivalent to eq. (\ref{IOC}).

\bibliography{C:/Users/cyndaquilka/OneDrive/Fizyka/bibliography}
\bibliographystyle{C:/Users/cyndaquilka/OneDrive/Fizyka/beztytulow2}

\end{document}